\documentclass{ws-procs9x6}

\def\etal {{\it et al.}}

\def\abw{\alpha(\overline{a}^w_{\rm eff})}
\def\abe{(\overline{a}^e_{\rm eff})}
\def\abp{(\overline{a}^p_{\rm eff})}
\def\abn{(\overline{a}^n_{\rm eff})}
\def\absun{(\overline{a}^\odot_{\rm eff})}
\def\cbw{(\overline{c}^w)}
\def\cbn{(\overline{c}^n)}
\def\cbp{(\overline{c}^p)}
\def\cbe{(\overline{c}^e)}

\def\lsim{\mathrel{\rlap{\lower4pt\hbox{\hskip1pt$\sim$}}
    \raise1pt\hbox{$<$}}}

\begin{document}

\title{STATUS OF
MATTER-GRAVITY COUPLINGS IN THE SME}

\author{JAY D.\ TASSON}

\address{Physics and Astronomy Department, Carleton College\\
Northfield, MN 55057, USA\\
E-mail: jtasson@carleton.edu}

\begin{abstract}
Constraints on Lorentz violation in matter-gravity couplings
are summarized along with existing proposals
to obtain sensitivities
that exceed current limits by up to 11 orders of magnitude.
\end{abstract}

\bodymatter

\section{Introduction}
The phenomenology of matter-gravity couplings and Lorentz violation
in the Standard-Model Extension (SME)\cite{foundation}
was developed in Refs.\ \refcite{akjtprl,akjt}.
The first portion of Ref.\ \refcite{akjt}
built upon existing analysis of the pure-gravity sector of the SME
\cite{lvpn}
to develop the necessary theoretical tools
for the experimental analysis.
These theoretical developments are summarized
in Ref.\ \refcite{jtcpt10}.
The second portion of Ref.\ \refcite{akjt}
generates explicit predictions for 
the detection of matter-sector coefficients for Lorentz violation
in a large number of gravitational tests.
Note that `explicit prediction' here means
that the experimental observable 
has been calculated and decomposed by signal frequency
such that it is ready to fit with experimental data.
Of special interest in this context
are sensitivities to the 12 coefficient components of $\abw_\mu$,
where $w$ runs over species, proton, neutron, and electron,
which are unobservable in the absence of gravity.\cite{akjtprl}
Experimental implications of the remaining spin-independent coefficient,
$\overline{c}_{\mu \nu}$
were also considered.
Consideration of spin-dependence in matter-gravity couplings
is now underway.\cite{yuri,ablt}
 
Predictions were made 
for the following tests:
laboratory tests
such as gravimeter experiments
and
tests of the Weak Equivalence Principle (WEP)
with ordinary neutral matter;
versions of these experiments with electrically charged matter,
higher-generation matter,
and antimatter;
WEP tests in space;
solar-system tests
such as lunar laser ranging and 
precession of the parihelion of various bodies;
and light-travel tests 
such as time-delay,
Doppler shift,
redshift,
and null-redshift tests.
These predictions were then used to place several constrains
on the relevant coefficients.
Following the publication of Refs.\ \refcite{akjtprl,akjt},
analysis of additional tests has been performed
obtaining additional constraints.
Section \ref{current} of this proceedings contribution
summarizes these existing limits.
The many proposed investigations that remain to be completed,
which could extend existing limits,
are summarized in 
Sec.\ \ref{todo}. 
Unless otherwise stated,
bounds on combinations of coefficient components
assume 
all other SME coefficients are zero.

\section{Constraints}
\label{current}

Concurrent with the development of experimental and observational predictions,
several constraints were placed using the published results of experiments.
Constraints were placed on 4 combinations
of the 12 $\abw_\mu$ coefficient components,
and 4 constraints were placed on previously unconstrained combinations
of $\cbw_{\mu \nu}$.\cite{akjt}
Four of these constraints are
from precession of the perihelion of bodes based on existing data, \cite{will}
\begin{eqnarray}
\nonumber
|-0.97 \absun_X 
 +0.15 \absun_Y 
 +0.18 \absun_Z| &\lsim& 10^{-6} \ {\rm GeV} \ {\rm (Mercury)},\\
\nonumber
|-0.97 \cbn_{TX} + 0.15 \cbn_{TY} + 0.18 \cbn_{TZ} | &\lsim& 10^{-5}  \ \phantom{\rm GeV} \ {\rm (Mercury)},\\
\nonumber
|-0.97 \absun_X 
\nonumber
 -0.21 \absun_Y
\nonumber
 -0.10 \absun_Z | &\lsim& 10^{-6} \ {\rm GeV}  \ {\rm (Earth)},\\
|-0.97 \cbn_{TX} - 0.21 \cbn_{TY} - 0.10 \cbn_{TZ} | &\lsim& 10^{-5}   \ \phantom{\rm GeV} \ {\rm (Earth)},
\end{eqnarray}
where $\absun_J = \alpha [\abe_J + \abp_J + 0.1 \abn_J]$.
Note that the $\cbn_{\Xi \Sigma}$ constraints above
are simplified using
existing constraints \cite{data} on other combinations of $\cbw_{\Xi \Sigma}$.
The other 2 constraints on $\abw_\Xi$ are from torsion pendulum measurements of WEP
based on data from Ref.\ \refcite{tpend},
\begin{equation}
\label{eq:at}
|\alpha \abn_T| \lsim 10^{-10} \ {\rm GeV},
\qquad
|\alpha \abe_T + \alpha \abp_T| \lsim 10^{-10} \ {\rm GeV},
\end{equation}
and the final constraints are based on combined results
from torsion pendulum WEP measurements and falling corner-cube WEP measurements \cite{fallingwep}
\begin{equation}
\label{eq:ctq}
|\cbn_Q| \lsim 10^{-8},
\qquad
|\cbe_{TT} + \cbp_{TT} - \cbn_{TT}| \lsim 10^{-8},
\end{equation}
though other WEP tests
could be used if sufficient sensitivity is available.

Following Refs.\ \refcite{akjt,akjtprl},
weak constraints
have been achieved 
on 4 additional combinations of $\abw_\mu$ coefficients:
\begin{eqnarray}
\nonumber
|\alpha \abn_X + 0.83 \alpha \left[ \abp_X + \abe_X \right]| &\leq& 0.2 \ {\rm GeV},\\
|\alpha \abn_Y + 0.83 \alpha \left[ \abp_Y + \abe_Y \right]| &\leq& 0.2 \ {\rm GeV},
\end{eqnarray}
using a torsion-strip balance, \cite{torsionstrip}
and 
\begin{eqnarray}
\nonumber
|\alpha \left[ \abe_X + \abp_X + \abe_X \right]| &=& 0.44 \pm 0.28 \ {\rm GeV},\\
|\alpha \left[ \abe_Y + \abp_Y + \abe_Y \right]| &=& 0.04 \pm 0.24 \ {\rm GeV},
\end{eqnarray}
via a reinterpretation \cite{gmag} of He/K comagnetometer $b_\mu$ results. \cite{hek}
Computational work on $\abw_J$ has also been done
associated with the Cassini mission.\cite{hees}
Additional work has been done
associated with the separation of $\abw_T$ and $\cbw_{TT}$ as well
resulting in the independent constraints \cite{hohensee} on
$\cbp_{TT}$ and $\cbn_{TT}$ at the $10^{-6}$ level,
$\alpha [\abe_T + \abp_T]$ and $\abn_T$ at the level of $10^{-6}$ GeV,
and mush stronger constraints on $\cbe_{TT}$ via nongravitational experiments.
Ref.\ \refcite{data}
summarizes all constraints discussed.

\section{Outstanding proposed analysis}
\label{todo}

To date, 8 combinations of the 12 components of the $\abw_\mu$ coefficient
have been constrained,
and 4 of those constraints are weak.
Existing experiments could improve the weak constraints
by up to 6 orders of magnitude.
Proposed experiments could improve these constraints
by up to 11 orders of magnitude,
and some could gain sensitivity to unconstrained combinations.

Analysis of existing data from the following experiments
could provide up to the indicated order of magnitude improvement
in sensitivities to spatial components of $\abw_\Xi$:
torsion pendulum WEP \cite{tpend}, 6 orders of magnitude;
super-conducting gravimeters \cite{super}, 6 orders;
lunar laser ranging \cite{llr}, 5 orders;
Cassini data \cite{hees}, 5 orders;
and atom interferometry \cite{fallingwep,hatom}, 4 orders.
If performed,
the following proposed experiments could provide even
greater improvement as indicated:
space-based WEP \cite{space}, up to 11 orders;
Earth-based WEP \cite{ewep,dghk}, up to 10 orders;
and gravimeters \cite{dghk}, up to 9 orders.

Gravitational tests with special types of matter
could obtain sensitivity to additional unconstrained combinations.
The 4 unconstrained combinations of $\abw_\Xi$ components
for ordinary matter could be accessed with charged-matter tests. \cite{charge}
Tests with higher-generation matter \cite{hiergen} of type $w$ would attain sensitivity to
many unconstrained associated coefficient components
of $\abw_\Xi$.
Tests with antimatter \cite{anti} also have the ability to separate special combinations \cite{akjt}
of coefficients if sufficient sensitivity can be reached.

The above possibilities offer excellent prospects for improved tests
of Lorentz symmetry,
and provide the opportunity for significant progress
in the ongoing search for new physics at the Planck scale.

\end{document}